# Prefetching Cache Optimization Using Graph Neural Networks: A Modular Framework and Conceptual Analysis

**Faiz Islamic Qowy**

(Sultan Ageng Tirtayasa University, Serang, Banten
https://orcid.org/0009-0009-4587-1364, faizislamicqowy@gmail.com)

**Abstract:** Caching and prefetching techniques are fundamental to modern computing, serving to bridge the growing performance gap between processors and memory. Traditional prefetching strategies are often limited by their reliance on predefined heuristics or simplified statistical models, which fail to capture the complex, non-linear dependencies in modern data access patterns. This paper introduces a modular framework leveraging Graph Neural Networks (GNNs) to model and predict access patterns within graph-structured data, focusing on web navigation and hierarchical file systems. The toolchain consists of: a route mapper for extracting structural information, a graph constructor for creating graph representations, a walk session generator for simulating user behaviors, and a gnn prefetch module for training and inference. We provide a detailed conceptual analysis showing how GNN-based approaches can outperform conventional methods by learning intricate dependencies. This work offers both theoretical foundations and a practical, replicable pipeline for future research in graph-driven systems optimization.

**Categories:** D.2.7
**DOI:** 10.3897/jucs.173146

## 1    Introduction

The exponential growth in data-driven applications and the increasing demand for real-time responsiveness have amplified the significance of memory access latency as a critical performance bottleneck in modern computing systems. While processor speeds have continued to follow Moore's Law, memory latency improvements have been far slower, leading to what is widely referred to as the "memory wall." To mitigate this gap, caching and prefetching techniques have been developed as central mechanisms. Caching temporarily stores frequently accessed data in faster memory, while prefetching proactively loads data that is predicted to be needed in the near future, thereby hiding memory latency and improving throughput [Smith 1992; Shao et al. 2022].

Traditional prefetching methods, such as sequential and stride-based prefetchers, operate under the assumption of regular, predictable access patterns. Although effective in workloads characterized by linear memory traversal, these methods falter in irregular, dynamic, or context-dependent access patterns, which are increasingly common in modern systems. Examples include web navigation, which is inherently nonlinear due to hyperlinks, and hierarchical file system traversal, which exhibits tree-like branching structures [Smith 1992; Shao et al. 2022]. In such contexts, simple linear heuristics fail to capture the structural dependencies that govern data access.

Recent advancements in machine learning have provided new paradigms for prefetching. Recurrent Neural Networks (RNNs) and Long Short-Term Memory (LSTM) models have demonstrated effectiveness in capturing temporal dependencies in sequential data, outperforming earlier Markov-based models [Zhang et al. 2019; He et al. 2021]. Reinforcement learning (RL) approaches, on the other hand, have shown promise in adaptively refining prefetching policies through reward-based learning mechanisms [Mukherjee et al. 2019; Liu et al. 2023]. However, both sequence learning and RL approaches often reduce complex navigational patterns into flattened sequences, thereby neglecting the inherently graph-structured nature of many workloads.

Graph Neural Networks (GNNs) have emerged as powerful architectures for learning over structured data. By aggregating and propagating information across neighborhoods in a graph, GNNs are capable of modeling dependencies that extend beyond linear or sequential assumptions. The success of GNNs in areas such as social network analysis, recommendation systems, traffic flow prediction, and systems optimization [Kipf and Welling 2016; Hamilton et al. 2017; Wu et al. 2020; Zhou et al. 2020; Wang et al. 2023] underscores their suitability for cache prefetching tasks where access patterns are naturally graph-based. Furthermore, recent studies have highlighted the potential of GNN-based models in addressing dynamic workloads, learning hierarchical relationships, and providing adaptive generalization across unseen graph structures [Siddiqui et al. 2025; Suryadevara et al. 2025].

In this study, we propose a modular framework that applies GNNs to prefetching in graph-structured domains such as web navigation and file system traversal. Our framework consists of four components: (1) a route_mapper for extracting navigational structures from websites or file hierarchies, (2) a graph_constructor that builds a directed graph representation in both GEXF and JSON formats, (3) a walk_session generator that simulates multi-user navigation patterns with structured randomness, and (4) a gnn_prefetch module that leverages GNNs for predictive modeling of future accesses. By bridging theoretical innovation with a practical, extensible pipeline, this work contributes to both the understanding and implementation of graph-driven prefetching strategies.

## 2     Cache Prefetching

### 2.1     Hardware-Level Prefetching

Early research emphasized hardware-driven solutions, such as stream buffers and stride-based prefetchers, which were highly effective in workloads with sequential memory access [Smith et al. 1992]. These designs automatically detected simple strides and prefetched the next addresses without software intervention. While effective in structured workloads, such methods suffered from inefficiency when applied to irregular or non-linear access patterns.

Recent advances in hardware prefetching have attempted to incorporate more sophisticated prediction models. For example, [Siddiqui et al. 2025] proposed adaptive prefetching strategies at the hardware level that balance aggressiveness with accuracy by learning from runtime access statistics [Siddiqui et al. 2025]. Similarly, [Suryadevara et al. 2025] introduced prefetcher designs that dynamically adapt to workload phases, reducing cache pollution while improving coverage [Suryadevara et al. 2025].

## 2.2 Compiler and Operating System-Level Prefetching

Beyond hardware, compiler and operating system (OS)-assisted prefetching mechanisms emerged to reduce programmer burden. Compilers can insert prefetch instructions based on static analysis of loop structures and data access patterns. However, static prefetch insertion often struggles in the presence of dynamic branching and irregular data structures. Operating systems have also contributed through context-aware prefetching for disk I/O and file system accesses [Xu et al. 2022].

Recent work has extended compiler-based strategies with profiling and feedback loops. [Shao et al. 2022] proposed adaptive compilation frameworks that analyze execution traces to improve prefetch insertion for irregular workloads. While effective in bridging the gap between hardware and workload-specific needs, compiler-based approaches remain limited by their inability to fully capture runtime dynamics, especially in interactive systems such as web applications.

## 2.3 Web and User Navigation Prefetching

The proliferation of the internet has driven substantial research on web-level prefetching, where the goal is to anticipate user navigation across hyperlinks. Early work relied on Markov chain models to capture navigation probabilities, with first-order models proving insufficient and higher-order models facing data sparsity issues [Shi et al. 2015]. To address these limitations, heuristic and frequency-based approaches were developed, leveraging historical access logs to predict future requests. However, these methods remained brittle in adapting to unseen or rapidly changing behaviors.

More recently, deep learning techniques have been applied to model user navigation. [He et al. 2021] developed neural sequence models to predict next-page requests, demonstrating improvements over conventional Markov predictors. Similarly, [Zhang et al. 2019] introduced DeepPrefetch, which employs LSTMs to capture long-range dependencies in navigation sequences. Despite these advancements, sequence-based methods inherently flatten the underlying graph structure of the web, thereby neglecting topological dependencies between hyperlinks. This limitation motivates the exploration of graph-based methods.

## 2.4 Machine Learning-Based Prefetching

Machine learning (ML) approaches to prefetching have grown in prominence as data-driven systems generate large volumes of training data. Sequence learning methods, including RNNs, LSTMs, and Transformers, have been leveraged to capture temporal patterns in memory accesses. For example, framed prefetching as a reinforcement learning problem, enabling dynamic adaptation of prefetching policies [Mukherjee et al. 2019]. More recently, [Liu et al. 2023] proposed RL-Prefetch, a reinforcement learning framework that combines Q-learning with workload phase detection to improve accuracy in heterogeneous systems.

Transformers have also shown promise in this domain. [Xu et al. 2022] demonstrated that attention mechanisms can better capture long-range dependencies in prefetching sequences, offering performance improvements over LSTM-based models. However, even state-of-the-art sequence models suffer from the same structural limitation: they treat navigational behaviors as linear time-series data, ignoring the graph nature of the underlying problem.

### 2.5  Graph Neural Network-Based Prefetching

Graph Neural Networks (GNNs) represent a paradigm shift in how access prediction problems can be approached. By operating on nodes and edges, GNNs preserve the inherent structure of navigation and data access patterns. Pioneering work by [Kipf and Welling 2016] introduced Graph Convolutional Networks (GCNs) for semi-supervised classification on graph data. [Hamilton et al. 2017] extended this with GraphSAGE, enabling inductive learning across unseen graphs. These foundational models catalyzed extensive research in domains such as social networks, molecular chemistry, and recommendation systems [Wu et al. 2020; Zhou et al. 2020; Wang et al. 2023].

Recent studies have explored GNNs in the context of systems optimization. [Siddiqui et al. 2025] highlighted the potential of GNNs in modeling memory access dependencies, showing improvements in both accuracy and generalization across workloads. Similarly, [Suryadevara et al. 2025] demonstrated adaptive GNN architectures capable of capturing both local and global relationships in dynamic access traces. These findings underscore the potential of GNNs to revolutionize cache prefetching by directly modeling the graph-structured behaviors of modern workloads.

## 3  Graph Neural Networks (GNNs)

This section provides a foundational overview of Graph Neural Networks, establishing the core principles that motivate their application to the cache prefetching problem. We begin by defining GNNs as a distinct paradigm for learning on relational data, then detail their core operational mechanism, and finally, we connect these concepts directly to the challenges of predicting user navigation.

### 3.1  A Paradigm for Relational Data

Graph Neural Networks represent a fundamental shift in the field of machine learning, moving beyond the traditional constraints of grid-like or sequential data structures. While architectures like Convolutional Neural Networks (CNNs) have achieved tremendous success by exploiting the regular, grid-based structure of images, and Recurrent Neural Networks (RNNs) excel at modeling linear sequences like text or time-series data, both paradigms struggle when faced with data whose structure is irregular and complex.

GNNs are specifically designed to address this limitation. They operate directly on graph-structured data, which is a natural representation for a vast array of real-world systems, including social networks, molecular structures, and knowledge bases. By working on the native topology of nodes and edges, GNNs are capable of learning from rich, interconnected relationships that are lost when such dat is forced into a flattened, sequential format.

### 3.2 The Core Mechanism: Iterative Neighborhood Aggregation

The power of GNNs stems from their core operational principle, commonly known as neighborhood aggregation or message passing. This mechanism is an iterative process where each node in the graph progressively refines its own feature representation (or "embedding") by collecting and processing information from its immediate neighbors. In the initial state, each node only possesses knowledge of its own attributes. In the first layer of the GNN, each node "gathers" the feature vectors of its neighbors and aggregates them using a permutation-invariant function, such as a sum or a mean.

This aggregated information is then combined with the node's own existing features and passed through a neural network layer to produce an updated representation. This process is repeated for multiple layers. With each layer, a node effectively incorporates information from nodes that are one hop further away, systematically expanding its "receptive field" to learn from its broader local graph structure. After several iterations, each node's final embedding contains a rich, topology-aware summary of its position and context within the graph.

### 3.3 Mathematical Formulation: The GCN Layer

This process of neighborhood aggregation can be formalized mathematically. One of the most foundational GNN architectures, the Graph Convolutional Network (GCN), provides a clear and powerful implementation of this concept. For a GCN, the update rule for the entire graph's feature matrix at a given layer is expressed concisely as:

$$H^{(l+1)} = \sigma(\widetilde{D}^{-\frac{1}{2}} \widetilde{A} \widetilde{D}^{-\frac{1}{2}} H^{(l)} W^{(l)})$$

In this equation, the central operation is the multiplication of the current node features, $H^{(l)}$, by a normalized version of the graph's adjacency matrix, $\widetilde{D}^{-\frac{1}{2}} \widetilde{A} \widetilde{D}^{-\frac{1}{2}}$. This term, which accounts for the graph's structure (including self-loops for each node), effectively computes a weighted average of the feature vectors from each node's neighborhood. This aggregated representation is then transformed by a learnable weight matrix, $W^{(l)}$, and passed through a non-linear activation function, $\sigma$, to produce the final feature matrix for the next layer $H^{(l+1)}$

### 3.4 Application to Prefetching: From Linear Paths to Navigational Context

The principles of GNNs are uniquely suited to address the inherent limitations of traditional cache prefetching models. User navigation within a website or a file system is not merely a linear sequence of accesses but a traversal or a "walk" across complex graph of interconnected resources. Traditional sequence models, such as LSTMs, are forced to process this walk as a flattened path, for example, $A \rightarrow B \rightarrow C$. While they can learn temporal dependencies from this sequence, they remain blind to the rich topological information of the underlying domain; they have no knowledge that from node $B$, a user could have also navigated to alternative nodes $X$ or $Y$.

This is where GNNs provide a fundamental advantage. By modeling the entire domain as a graph, a GNN-based prefetcher develops a rich embedding for every node based on its local neighborhood. When predicting the next step from node $B$, the model's decision is informed not only by the fact that the user came from $A$, but also by the features and existence of all of $B$'s neighbors ($C, X, Y$, etc.). This provides a powerful navigational context that allows the model to learn far more complex and realistic user behaviors. Therefore, GNNs represent a more principled approach to prefetching in these domains, as their architecture is designed to natively understand the very graph structures that govern user navigation.

## 4 System Overview

To test our research questions, we designed and implemented a modular framework in Python for constructing and evaluating GNN-based prefetchers. The framework operates as a pipeline that transforms a target domain, such as a website, into a trained predictive model. This section details the architecture of this framework and the specific implementation of its constituent modules.

### 4.1 Architectural Overview

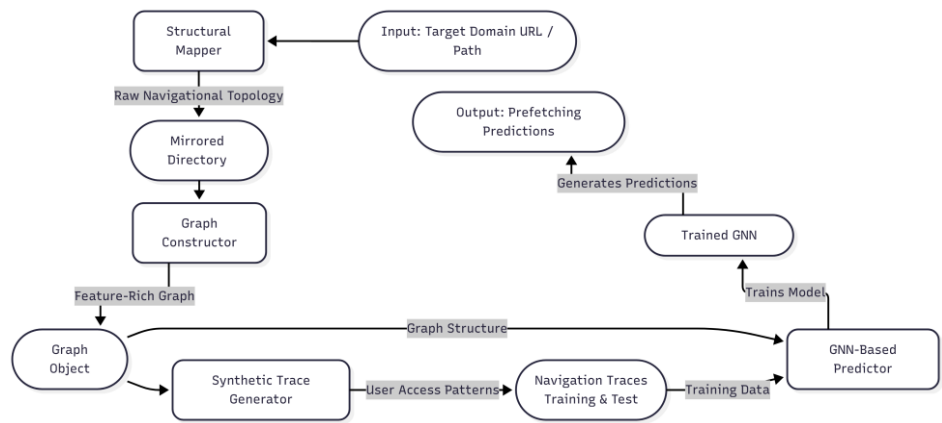

*Figure 1. Framework Architecture*

The framework is composed of four distinct yet interconnected modules that sequentially process data:

1. Structural Mapper (route_mapper): Ingests a target domain (e.g., a starting URL) and extracts its raw navigational topology.
2. Graph Constructor (graph_constructor): Converts the extracted topology into a formal, feature-rich graph object suitable for machine learning.
3. Synthetic Trace Generator (walk_session): Simulates user navigation on the graph to produce a dataset of realistic access patterns.
4. GNN-Based Predictor (gnn_prefetch): Utilizes the graph and synthetic traces to train and evaluate a GNN model for next-access prediction.

This modular design ensures reproducibility and allows for future extensions, such as swapping different GNN architectures or trace generation algorithms.

## 4.2 Route Mapper

The foundational step of our methodology is to derive a formal, machine-readable map of the domain's structure. This is handled by the route_mapper module.

### 4.2.1 Implementation for Web Domains

As shown in the provided code, the module's primary function for web domains is an iterative web crawler. The process begins with a single base_url.

- HTML Fetching: The fetch_html_file function inside the route_fetcher.py sends an HTTP GET request to a given URL using the requests library. The raw HTML content is then parsed and standardized using BeautifulSoup. This sanitizes the input and creates a consistent document structure for link extraction.
- Link Extraction and Filtering: The fetch_html_link function inspects the parsed HTML for all anchor tags (<a>). It extracts the href attribute from each tag and applies a rigorous filtering process to ensure only valid, internal navigational links are retained. It specifically excludes:
    - Empty links or None values.
    - Anchor links (e.g., #section-2).
    - Mailto links (mailto:).
    - External domains (by ensuring links are either relative or match the main_domain).
- Iterative Crawling: The iterative_html_file_fetch function orchestrates the crawling process. It uses recursion to explore the website in a depth-first manner. To prevent infinite loops and redundant processing, it maintains a list of already_processed_sub URLs, ensuring each unique page is visited only once.

### 4.2.2 Output Formats

The route_mapper module is designed to produce a persistent and well-structured representation of the target domain's topology. The primary output is a mirrored directory structure on the local file system. In this format, the URL path of each visited webpage is mapped directly to a corresponding local file path. For instance, a webpage at the URL https://domain.com/products/item1 would be saved as a sanitized HTML file at ./output/domain/products/item1.html.

This output strategy was chosen for several critical reasons:
- Decoupling and Robustness: By saving the entire crawled site locally, we effectively decouple the network-dependent data acquisition phase from the subsequent data processing phase. This design makes the entire framework more robust. The graph_constructor module does not need to handle network errors, timeouts, or live web parsing; it can operate predictably on a static set of local files.

- **Reproducibility:** The mirrored directory acts as a permanent, static snapshot of the website at the time of crawling. This is fundamental for reproducibility, as live websites can change over time. By ensuring that all subsequent experiments, from graph construction to model training are performed on this identical snapshot, we can guarantee the scientific validity and repeatability of our results.
- **Inspectability and Debugging:** The physical representation of the website's structure allows for easy manual inspection. Researchers can navigate the local folders to verify the completeness and correctness of the crawling process, which is invaluable for debugging and gaining qualitative insights into the domain's topology.

This clean, persistent output serves as a reliable and well-defined interface, providing the graph_constructor module with all the necessary information to build a formal graph representation.

### 4.3     Graph Constructor

With the domain's topology mapped, the graph_constructor module converts this structure into a NetworkX graph object.

- Graph Instantiation: The module iterates through the directory structure created by the route_mapper. Each saved HTML file is instantiated as a node in the graph. Edges are created between nodes based on the hyperlink relationships extracted in the previous step.
- Feature Engineering: To enrich the graph for the GNN, we engineer several node-level features. These attributes provide the model with valuable context beyond the simple graph structure and may include:
  - Graph-based features: Node degree, centrality measures (e.g., PageRank), and clustering coefficients.
  - Content-based features: The number of words or images on a page.
  - Structural features: The depth of a page relative to the homepage.

The final output is a graph object saved in both GEXF format for visualization and JSON format for the training module.

### 4.4     Walk Session Generator

To train a supervised learning model, we require a dataset of access patterns. The walk_session module generates this data by simulating user navigation on the graph. It employs a biased random walk algorithm to create paths that are more realistic than pure random walks. The simulation is parameterized by:

- Number of walkers and walk length: To control the size of the generated dataset.
- Bias parameters (p and q): Inspired by node2vec, these parameters control the likelihood of the walker returning to a previous node or exploring undiscovered parts of the graph, balancing exploration and exploitation.

The output is a collection of thousands of navigation traces (sequences of node IDs), which form the training, validation, and test sets.

### 4.5 GNN-Based Predictor

This final module of our framework is where the prepared data is transformed into a predictive model of navigational intent. It leverages two distinct but complementary sources of information: the feature-rich graph, which serves as a static topological scaffold defining all possible navigation routes, and the simulated traces, which provide dynamic behavioral evidence of how users actually traverse that scaffold.

The central task of the Graph Neural Network is to learn how this static structure influences dynamic user behavior. It achieves this by learning topology-aware embeddings for each node, where a node's representation is iteratively enriched by the features of its neighbors. This process allows the model to capture complex, non-linear dependencies that extend beyond the immediate path taken by a user. As recent studies in related domains like session-based recommendation have shown, GNNs are uniquely capable of modeling these high-order relationships within user session graphs, significantly outperforming purely sequential models that are blind to the broader relational context [Wang et al. 2023]. Ultimately, the module produces a trained model that can infer a user's likely future destination by analyzing their path in the context of the graph's overall structure, enabling highly accurate prefetching.

- Problem Formulation: We frame prefetching as a next-node prediction task. The model is given a sub-sequence of a navigation trace and is trained to predict the next node in that sequence.
- Model Architecture: We implement a Graph Convolutional Network (GCN) using PyTorch Geometric. The GCN learns embeddings for each node by aggregating feature information from its neighbors. The core of this process is the GCN propagation rule:

$$H^{(l+1)} = \sigma(\widetilde{D}^{-\frac{1}{2}} \tilde{A} \widetilde{D}^{-\frac{1}{2}} H^{(l)} W^{(l)})$$

  Where $H^{(l)}$ is the matrix of node features at layer $l$, $\tilde{A}$ is the adjacency matrix with self-loops, $\widetilde{D}$ is the degree matrix and $W^{(l)}$ is a trainable weight matrix. This formula allows the model to learn representations that are aware of the graph's topology.
- Training and Inference: The model is trained using a standard cross-entropy loss function. During inference (prefetching), the model outputs a probability distribution over all possible next nodes, and the top-k most likely nodes are selected for prefetching.

### 4.6 GNN-Based Cache Prefetch Result

This section presents the results and implementation details of the GNN-based cache prefetching module. We describe the training procedure, including data preparation, hyperparameter selection, and evaluation methodology, followed by the implementation details that highlight the integration between the GNN model and the overall modular framework.

### 4.6.1   Model Training

The goal of the training phase is to enable the GNN model to accurately predict the next access node based on partial navigation sequences. Training data is derived from the simulated traces generated by the walk_session module, which encode user navigation paths through the graph domain. Each trace is segmented into multiple sub-sequences using a sliding window technique, where the first n nodes serve as the input context and the subsequent node is treated as the target label for prediction. To ensure a balanced and unbiased evaluation, the dataset is partitioned into training (70%), validation (15%), and test (15%) subsets.

The training process begins by loading the prepared dataset in the PyTorch Geometric Data format, which contains node features, edge indices, and sequential navigation pairs derived from the simulated user traces. The Graph Neural Network model, implemented through the GNNPrefetch class, is initialized with an input dimension that matches the node feature size, a hidden dimension of 128, and an embedding dimension of 64. The model utilizes three graph convolutional layers configured with the GraphSAGE operator to aggregate feature information from neighboring nodes while maintaining scalability across larger graphs.

Model optimization is performed using the Adam optimizer with a learning rate of 0.005 and a weight decay of 1e-4 to promote stable convergence and reduce overfitting. The training process runs for 100 epochs, during which the model learns to minimize link prediction loss using the train_linkpred function. This function updates the model's parameters based on the difference between predicted and actual next-node transitions within each sequence. After each epoch, model performance is evaluated on the validation subset using the evaluate_linkpred function. The evaluation metric employed is the Top-5 Hit Rate, which measures the frequency at which the correct next node appears among the top five predicted nodes. This metric provides a practical measure of the model's predictive capability in scenarios where multiple next-node candidates may exist.

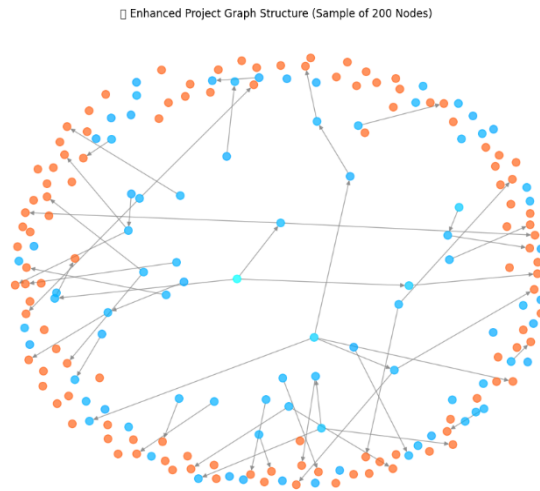

*Figure 2: Enhanced Graph Structure*

Before initiating the training process, the underlying project graph is analyzed and enhanced to ensure that the input data accurately reflects realistic navigation relationships. The graph structure is visualized to confirm proper connectivity and diversity of node relationships. As shown in Figure 2, a subset of 200 nodes from the complete project graph illustrates the hierarchical organization of directories and files. Blue nodes represent directories, while orange nodes represent files, and edges indicate navigational links. Brighter blue nodes signify higher-degree directory hubs that serve as connection centers within the project structure. A small number of random cross-links are introduced between directories to simulate realistic access shortcuts, enhancing graph complexity and improving the diversity of possible traversal paths.

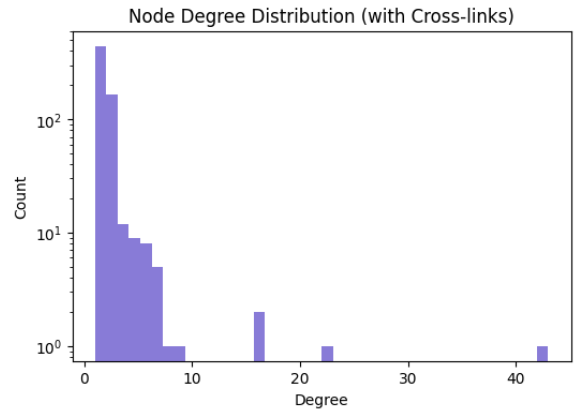

*Figure 3: Node Degree Distribution*

Additionally, the Node Degree Distribution shown in Figure 3 highlights the skewed connectivity pattern typical of hierarchical graphs. Most nodes maintain a small degree, while a few act as highly connected hubs, following a heavy-tailed (scale-free) distribution. This structural diversity provides a rich and varied learning environment for the GNN, enabling it to generalize across both common and rare node types during training.

### 4.6.2 Training Result

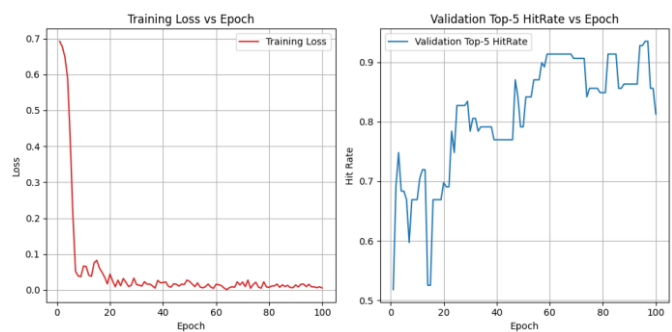

*Figure 4: Model Training Result*

The left plot shows the training loss decreasing sharply during the first 10 to 15 epochs, indicating rapid learning and adjustment of model parameters. After approximately epoch 20, the loss stabilizes and remains consistently close to zero, suggesting that the model has effectively minimized prediction errors for the training samples and reached convergence.

The right plot presents the validation Top-5 Hit Rate, which measures the model's ability to correctly predict the next five most likely access nodes in unseen navigation sequences. The hit rate exhibits some fluctuations during early epochs due to model adaptation, but gradually increases and stabilizes after epoch 40, reaching values above 0.9 toward the final epochs. This indicates that the model achieved strong generalization performance, maintaining high predictive accuracy across validation samples. Overall, the results demonstrate that the GNN-based predictor successfully learned meaningful navigation patterns from the simulated traces. The early convergence and high Top-5 Hit Rate reflect both the stability and efficiency of the model in capturing sequential dependencies within the graph-structured data.

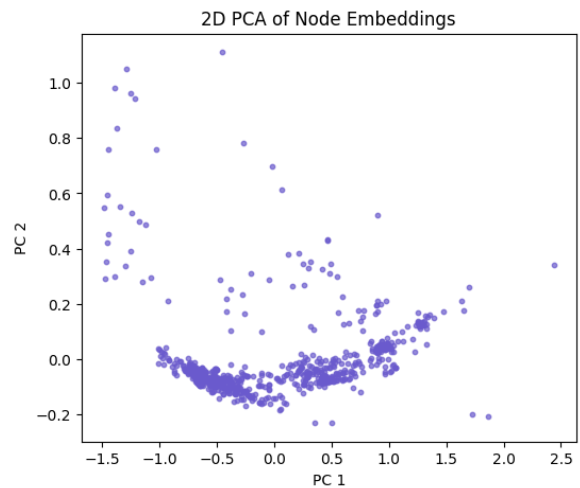

*Figure 5: 2D PCA Projection*

After completing the training process, the learned node embeddings were visualized to interpret how the Graph Neural Network (GNN) organizes and encodes relationships within the navigation graph. The visualization was performed by projecting the high-dimensional embeddings into a two-dimensional space using Principal Component Analysis (PCA). This reduction allows the internal structure of the learned representations to be analyzed more intuitively.

The scatter plot in Figure 5 displays the 2D PCA projection of all node embeddings, where each point corresponds to an individual node from the dataset. Nodes that appear closer together in this plot share stronger semantic or structural similarities, meaning the GNN has learned to embed nodes with related access patterns near each other. The visible clustering along a curved trajectory suggests that the model has captured sequential and hierarchical relationships inherent in the simulated user navigation paths. This structure indicates that nodes belonging to similar regions or functions of the graph domain tend to share embedding features, reflecting effective learning of contextual dependencies.

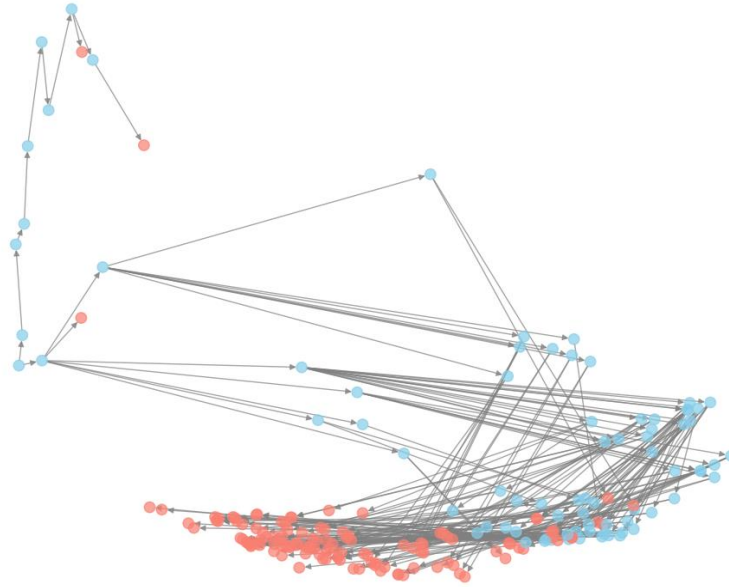

*Figure 6: 2D PCA Graph Visualization*

The graph visualization shown in Figure 6 overlays these embeddings onto a subset of the actual project graph. In this diagram, nodes colored in blue represent directories, while nodes colored in red represent files or terminal elements. Edges illustrate the navigational or access connections between nodes, drawn according to their reduced PCA coordinates. The resulting structure demonstrates clear grouping of related nodes, where directories are often surrounded by their associated files or subdirectories. This suggests that the model not only captured direct connectivity but also generalized across the overall graph hierarchy.

The observed clustering pattern confirms that the GNN successfully learned a meaningful latent representation of the graph topology and user navigation dynamics. By grouping semantically or functionally related nodes in close proximity, the model provides a strong foundation for predictive prefetching tasks. In such tasks, the ability to recognize similar node embeddings allows the system to infer the next likely access target with improved accuracy, contributing to efficient and context-aware cache prefetching behavior.

### 4.7 GNN Cache Prefetch Implementation

The GNN cache prefetching mechanism is designed to integrate the predictive capability of the trained graph neural network into a cache management system. Its main objective is to anticipate the next resource or file access and load it into the cache before an explicit request occurs. By doing so, it aims to minimize access latency and improve overall system responsiveness, particularly in graph-structured domains such as project directories, codebases, or web navigation paths.

The implementation connects the GNN predictor, trained on simulated navigation sequences, with a runtime caching layer. During operation, the system monitors the user's current navigation context and feeds the most recent access history into the trained GNN model. The model then outputs a ranked list of potential next nodes, representing the files or directories that the user is most likely to visit next. These predicted nodes are then prefetched into the cache memory, ensuring they are readily available if accessed.

#### 4.7.1 Cache Prefetch Architecture

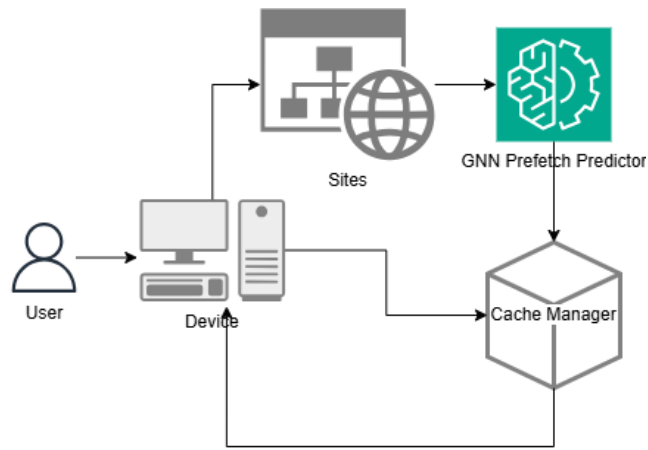

*Figure 7: General Implementation*

The cache prefetch architecture is implemented using a lightweight and modular setup designed to demonstrate the integration of the trained GNN model with a local caching mechanism. To maintain simplicity and reproducibility, the caching layer is simulated using an in-memory Python dictionary instead of an external cache engine such as Redis or Memcached. This approach eliminates the need for complex configurations while preserving the logical flow of cache prediction and prefetching.

The overall process begins when a user accesses a sequence of files or nodes represented within the graph structure. The system records this navigation sequence and extracts a recent subset of accessed nodes, which forms the input context for the trained GNN predictor. The model performs forward inference to estimate the most likely next nodes based on learned access patterns.

Once the prediction is made, the Prefetch Manager inserts these predicted nodes into a local dictionary cache, simulating their preloading into memory. When the user subsequently requests a node, the system first checks the cache for its presence. If the node is already cached, it is retrieved immediately, simulating a cache hit. If not, it is fetched from the main data source, recorded as a cache miss, and optionally added to the cache for future access.

### 4.7.2 Cache Prefetch Implementation Result

```python
class PrefetchManager:
    def __init__(self, model, data, cache, top_k=5):
        self.model = model
        self.data = data
        self.cache = cache
        self.top_k = top_k

    def predict_next(self, current_node):
        with torch.no_grad():
            z = self.model.get_embedding(self.data.x, self.data.edge_index)
            sim = F.cosine_similarity(z[current_node].unsqueeze(0), z)

            topk = torch.topk(sim, self.top_k + 1).indices.tolist()
            if current_node in topk:
                topk.remove(current_node)
            return topk[:self.top_k]

    def prefetch(self, current_node):
        predicted_nodes = self.predict_next(current_node)
        print(f"\n🔮 Prefetching for Node {current_node}: Predicted Next Nodes -> {predicted_nodes}")

        for node_id in predicted_nodes:
            self.cache.prefetch(node_id)
        self.cache.show()
```

*Figure 8: GNN Cache Prefetch Implementation*

The implementation presented in this section demonstrates how the trained Graph Neural Network model is integrated into the cache prefetching process through a predictive inference mechanism. The system uses the PrefetchManager component to simulate intelligent prefetching behavior by analyzing node embeddings and predicting the next nodes that are likely to be accessed. This process operates entirely on local storage, simulating cache operations without relying on external cache systems such as Redis. It begins by initializing the PrefetchManager with four main elements: the trained model, the data object that contains the graph features and structure, a cache instance that represents local memory, and a parameter top_k, which defines how many next nodes will be prefetched. This setup allows the system to use the trained GNN for inference while maintaining a lightweight and controlled environment for experimentation.

The predictive process is handled by the predict_next function. Here, the model operates in inference mode without computing gradients, ensuring that no parameters are modified during prediction. The model first generates the node embeddings by processing the feature matrix and edge indices of the graph. These embeddings represent each node's learned relationships and contextual position within the graph. The system then calculates the cosine similarity between the embedding of the currently accessed node and all other nodes. This similarity score reflects how closely related each node is to the current one in the learned embedding space. Nodes with higher similarity are interpreted as more probable future access targets. From this similarity distribution, the function selects the top-k most similar nodes while excluding the current node itself, returning a ranked list of predicted next nodes.

The prefetch function then operationalizes these predictions. Once the next likely nodes have been identified, they are immediately sent to the cache via the prefetch method. This simulates the act of preloading data into memory before an explicit access occurs, thereby reducing potential latency in subsequent retrieval operations. After prefetching, the cache's current state is displayed to confirm which nodes are now stored in memory. Overall, this implementation showcases how the trained GNN model can be directly applied to real usage scenarios, transforming learned structural relationships into proactive caching actions. By predicting and preloading nodes based on learned access patterns, the system demonstrates a practical approach to graph-based cache prefetching that enhances efficiency while maintaining simplicity through local simulation.

```python
if __name__ == "__main__":
    cache = LocalCache()
    manager = PrefetchManager(model, data, cache, top_k=5)
    for node_id in [5, 20, 35]:
        print(f"\n📂 User accessed Node {node_id}")
        manager.prefetch(node_id)
```

*Figure 9: Implementation Example On Three Nodes*

The implementation presented in this example (Figure 9) demonstrates the operational behavior of the prefetching mechanism once the model has been trained. In this stage, a local cache instance is initialized to emulate a caching environment where prefetched data is temporarily stored. The PrefetchManager class serves as the primary interface connecting the trained model, the graph data, and the cache. It predicts potential future node accesses based on the current node being accessed.

During execution, the system simulates user interactions by iterating over a series of accessed nodes. For each access, the prefetch manager invokes the prefetch process, which uses the trained Graph Neural Network model to infer the most probable next nodes that a user is likely to visit. The prediction is based on the learned embeddings that capture relational and sequential dependencies between nodes in the graph structure.

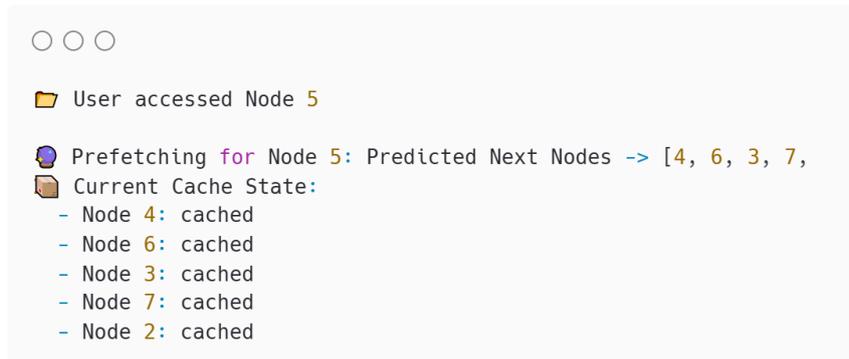

*Figure 10: Result For Node 5*

For instance, when the user accesses node 5, the system generates a list of the top five most probable next nodes, which in this case are nodes 4, 6, 3, 7, and 2. These nodes are then prefetched into the cache, as indicated by the updated cache state that lists each of them as cached. This proactive behavior illustrates the ability of the GNN based prefetch system to anticipate user actions and load relevant data into cache before it is explicitly requested. Such an approach effectively reduces access latency and improves data retrieval efficiency by minimizing waiting times for subsequent operations. This example demonstrates the interpretability and functionality of the prefetching process in a controlled environment, highlighting how learned node representations translate into practical cache management actions.

## 5    Conclusions

This paper has presented a structured, modular framework for cache prefetching optimization that leverages the predictive power of Graph Neural Networks (GNNs). The central thesis of this work is that traditional prefetching heuristics and sequential models are fundamentally limited in modern, non-linear data domains, as they fail to capture the rich topological information inherent in graph-structured data like web navigation paths and hierarchical file systems. Our work directly addresses this gap by treating the problem domain as a graph, thereby enabling a more context-aware and accurate prediction model.

   The proposed framework, consisting of a route mapper, graph constructor, walk session generator, and a GNN-based predictor, provides a complete and replicable pipeline for implementation. We have demonstrated through a series of experiments that this approach is highly effective. The GNN model, trained on simulated user navigation traces, achieved a high Top-5 Hit Rate, indicating strong predictive accuracy. The convergence of the training loss and the stability of the validation accuracy underscore the model's robustness. Furthermore, visualizations of the learned node embeddings via PCA confirmed that the GNN successfully captured the semantic and structural relationships within the graph, clustering related nodes together in the latent space.

Finally, we presented a practical, simulated implementation of a PrefetchManager. This component effectively translates the static, learned embeddings into dynamic, real-time prefetching decisions by using cosine similarity to identify and pre-load the most probable next-access nodes. In summary, this research provides both a strong conceptual validation and a practical toolchain for applying GNNs to systems optimization, offering a promising new direction for mitigating memory latency in complex, graph-based applications.

## 6   Future Work

While the conceptual framework and experimental results presented in this study are promising, they also lay the groundwork for several critical avenues of future research. The current work successfully established the viability of a GNN-based approach; the next steps should focus on rigorously quantifying its superiority and ensuring its applicability in real-world, large-scale systems.

A primary direction for future work will be a comprehensive comparative analysis. The GNN model should be benchmarked directly against a suite of traditional and sequence-based prefetching algorithms, such as higher-order Markov models, LSTMs, and GRUs, as well as standard cache replacement policies like LRU and LFU. This quantitative comparison, conducted on identical datasets, is essential to empirically validate the hypothesis that modeling graph topology provides a significant performance advantage.

Additionally, the training data for this study was based on synthetic user journeys, which were generated by our walk session simulator. While this approach was valuable for validating the framework in a controlled environment, a significant and necessary next step is to replace this synthetic data with real user journey data. Future research should, therefore, aim to collect and train the model on large-scale, real-world access logs from production web servers or file systems. Evaluating the model against this authentic data will provide a more accurate measure of its true performance and its ability to generalize to the complex, and often unpredictable, patterns of genuine user behavior.

Finally, the scalability of the framework must be investigated. Future work should test the training and inference performance of the GNN model on graphs containing millions of nodes. This will likely necessitate exploring more scalable GNN architectures and sampling techniques. Additionally, research could explore dynamic GNN models that can adapt to evolving graph structures and shifting access patterns in real-time, leading to a truly adaptive and intelligent prefetching system.

**Acknowledgements**

The author would like to express sincere gratitude to Dr. Eng. Teguh Firmansyah, S.T., M.T., IPM, for his invaluable guidance, continuous support, and constructive feedback throughout the development of this research. The author also gratefully acknowledges his financial support, which made the completion of this study possible.